\documentclass[aps,twocolumn,showpacs,preprintnumbers,amsmath,amssymb,
nofootinbib,showkeys,floatfix]{revtex4-1}

\usepackage{latexsym}
\usepackage{amsmath}
\usepackage{amssymb}

\usepackage{supertabular} 
\usepackage{placeins}
\usepackage{epsfig}
\usepackage{graphicx}
\usepackage{graphics}

\def\tstrut{\vrule height2.25ex depth0pt width0pt} 

\usepackage{hyperref} 
\hypersetup{
   colorlinks=true,
   linkcolor=blue,
   citecolor=blue,
   urlcolor=blue
}

\begin{document}

\title{Scaling of the $\mathbf{^{3}P_{0}}$ strength in heavy meson strong
decays}
\author{J. Segovia}
\author{D.R. Entem}
\author{F. Fern\'andez}
\affiliation{Grupo de F\'isica Nuclear and IUFFyM, \\ Universidad de 
Salamanca, E-37008 Salamanca, Spain}
\date{\today}

\begin{abstract}

The phenomenological $^{3}P_{0}$ decay model has been extensively applied to
calculate meson strong decays. The strength $\gamma$ of the decay interaction is
regarded as a free flavor independent constant and is fitted to the data. We
calculate through the $^{3}P_{0}$ model the total strong decay widths of the
mesons which belong to charmed, charmed-strange, hidden charm and hidden bottom
sectors. The wave function of the mesons involved in the strong decays
are given by a constituent quark model that describes well the meson
phenomenology from the light to the heavy quark sector. A global fit of the
experimental data shows that, contrarily to the usual wisdom, the $\gamma$
depends on the reduced mass of the quark-antiquark pair in the decaying meson. With
this scale-dependent strength $\gamma$, we are able to predict the decay width
of orbitally excited $B$ mesons not included in the fit.

\end{abstract}


\maketitle

\section{INTRODUCTION}
\label{sec:introduction}

Since the discovery in 1974 of the $J/\psi$ and, three years later, the $\Upsilon$
states, charmonia and bottomonia have been throughly studied, and still are a
subject of intensive theoretical and experimental research (see for instance
Ref.~\cite{springerlink:10.1140/epjc/s10052-010-1534-9}). The fundamental reason
is that a nonrelativistic picture seems to hold for them and they constitute the
simplest nontrivial system that can be used to test basic properties of QCD in
its nonperturbative regime.

In particular, the heavy meson spectra can be reasonably understood in
nonrelativistic models with simple or sophisticated versions of the
funnel potential, containing a short-range Coulomb-type term coming from
one-gluon exchange plus a long-range confining term.

However, meson strong decay is a complex nonperturbative process that has not
yet been described from first principles of QCD. This leads a rather poorly
understood area of hadronic physics which is a problem because decay widths
comprise a large portion of our knowledge of the strong interaction. 

Several phenomenological models have been developed to deal with this topic. The
most popular are the $^{3}P_{0}$
model~\cite{Micu:1968mk,PhysRevD.8.2223,PhysRevD.9.1415} and the flux-tube
model~\cite{PhysRevD.32.189,PhysRevD.35.907,PhysRevD.50.6855}. Both decay models
assume that a quark-antiquark pair is created with vacuum quantum numbers,
$J^{PC}=0^{++}$, but the flux-tube model includes the overlaps of the flux-tube
of the initial meson with those of the two outgoing mesons.

The $^{3}P_{0}$ model was first proposed by Micu~\cite{Micu:1968mk}. Le Yaouanc
{\it et al.} applied subsequently this model to meson~\cite{PhysRevD.8.2223} and
baryon~\cite{PhysRevD.9.1415} open-flavor strong decays in a series of
publications in the 1970s. They also evaluated strong decay partial widths of
the three charmonium states $\psi(3770)$, $\psi(4040)$ and $\psi(4415)$ within
the same model~\cite{LeYaouanc1977397,LeYaouanc197757}.

The $^{3}P_{0}$ model, which has since been applied extensively to the decays of
light mesons and baryons~\cite{summary3P0}, was originally adopted largely due
to its success in the prediction of the $D/S$ amplitude ratio in the decay
$b_{1} \to \omega\pi$. Another success of the decay model is that it predicts a
zero branching fraction ${\cal B}(\pi_{2}(1670)\to b_{1}\pi)$ and it is
experimentally measured to be $<1.9\times10^{-3}$ at $97.7\%$ confidence level.
It would not necessary be negligible in a different decay model.

An important characteristic, apart from its simplicity, is that the model
provides the gross features of various transitions with only one parameter, the
strength $\gamma$ of the decay interaction, which is regarded as a free
constant and is fitted to the data. It is generally believed that the
pair-production strength parameter $\gamma$, is roughly flavor-independent for
decays involving production of $u\bar{u}$, $d\bar{d}$ and $s\bar{s}$ pairs. As
an example, one can mentioned the work of Ref.~\cite{PhysRevD.72.094004} where a
total of $32$ experimentally well-determined decay rates have been fitted using
the $^{3}P_{0}$ model. The large experimental errors preclude definitive
conclusions about the dependence of $\gamma$ with respect the flavor sector and
the authors followed the convention of using a unique value for the $\gamma$
parameter. However, it is important to note that only $3$ of the total $32$
decay modes are referred to the heavy quark sector. They are $D^{\ast+}\to
D^{0}\pi^{+}$, $\psi(3770)\to D\bar{D}$ and $D_{s2}^{\ast}\to
DK+D^{\ast}K+D_{s}\eta$. Strong decay widths of mesons containing $b$-quark are
not treated and the remaining $29$ decay modes involve light and strange mesons.

Some attempts have been done to find a possible dependence of the vertex
parameter $\gamma$. In particular, Bonnaz and Silvestre-Brac has studied $10$
different $p$ dependences of the $\gamma$ parameter, where $p$ is the relative
momentum of the created $q\bar{q}$ pair. The model was only applied to mesons
involving light quarks. Although some improvement in the description of the
data has been found, it depends very crucially from the vertex form which is
arbitrary and unconstrained.

Our purpose here is to find a scale dependence of $\gamma$ from the light to the
heavy quark sector using a fit to the decay widths of the mesons which belong
to charmed, charmed-strange, hidden charm and hidden bottom sectors calculated
with the $^{3}P_{0}$ model. Certainly, the theoretical results have some
uncertainties coming from the decay model itself in the description of the
creation vertex and the wave functions used. Therefore, we expect to reach a
global description of the meson strong decays in every sector, not to take into
account the details of each decay mode. 

The wave functions for the mesons involved in the open-flavor strong decays are
the solutions of the Schr\"odinger equation with the potential model described
in Ref.~\cite{vijande2005constituent} using the Gaussian Expansion
Method~\cite{Hiyama:2003cu}. The model has recently been applied to mesons
containing heavy quarks in
Refs.~\cite{PhysRevD.78.114033,PhysRevD.80.054017,segovia2011semileptonic},
where different properties as spectra, strong decays and weak decays, has been
successfully explained.

In the paper we proceed as follows. In Sec.~\ref{sec:3P0model} we review the
$^{3}P_{0}$ decay model adapted to our formalism. Sec.~\ref{sec:running} is
devoted to the parametrization of the the strength $\gamma$ of the decay
interaction as a function of the scale. In Sec.~\ref{sec:results} we present
our results, comments about them are also included. Finally, we give some
remarks and conclusions in Sec.~\ref{sec:conclusions}.

\section{THE $\mathbf{^{3}P_{0}}$ DECAY MODEL}
\label{sec:3P0model}

\subsection{Transition operator}

The interaction Hamiltonian involving Dirac quark fields that describes the
production process is given by
\begin{equation}
H_{I}=\sqrt{3}\,g_{s}\int d^{3}x \, \bar{\psi}(\vec{x})\psi(\vec{x}),
\label{eq:IH3P0}
\end{equation}
where we have introduced for convenience the numerical factor $\sqrt{3}$, which
will be canceled with the color factor.

If we write the Dirac fields in second quantization and keep only the
contribution of the interaction Hamiltonian which creates a $(\mu\nu)$
quark-antiquark pair, we arrive, after a nonrelativistic reduction, to the
following expression for the transition operator
\begin{equation}
\begin{split}
T =& -\sqrt{3} \, \sum_{\mu,\nu}\int d^{3}\!p_{\mu}d^{3}\!p_{\nu}
\delta^{(3)}(\vec{p}_{\mu}+\vec{p}_{\nu})\frac{g_{s}}{2m_{\mu}}\sqrt{2^{5}\pi}
\,\times \\
&
\times \left[\mathcal{Y}_{1}\left(\frac{\vec{p}_{\mu}-\vec{p}_{\nu}}{2}
\right)\otimes\left(\frac{1}{2}\frac{1}{2}\right)1\right]_{0}a^{\dagger}_{\mu}
(\vec{p}_{\mu})b^{\dagger}_{\nu}(\vec{p}_{\nu}),
\label{eq:Otransition2}
\end{split}
\end{equation}
where $\mu$ $(\nu)$ are the spin, flavor and color quantum numbers of the
created quark (antiquark). The spin of the quark and antiquark is coupled to
one. The ${\cal Y}_{lm}(\vec{p}\,)=p^{l}Y_{lm}(\hat{p})$ is the solid harmonic
defined in function of the spherical harmonic.

As in Ref.~\cite{PhysRevD.54.6811}, we fix the relation of $g_{s}$ with the
dimensionless constant giving the strength of the quark-antiquark pair creation
from the vacuum as $\gamma=g_{s}/2m$, being $m$ the mass of the created quark
(antiquark).

\subsection{Transition amplitude}

\begin{figure}[!t]
\begin{center}
\epsfig{figure=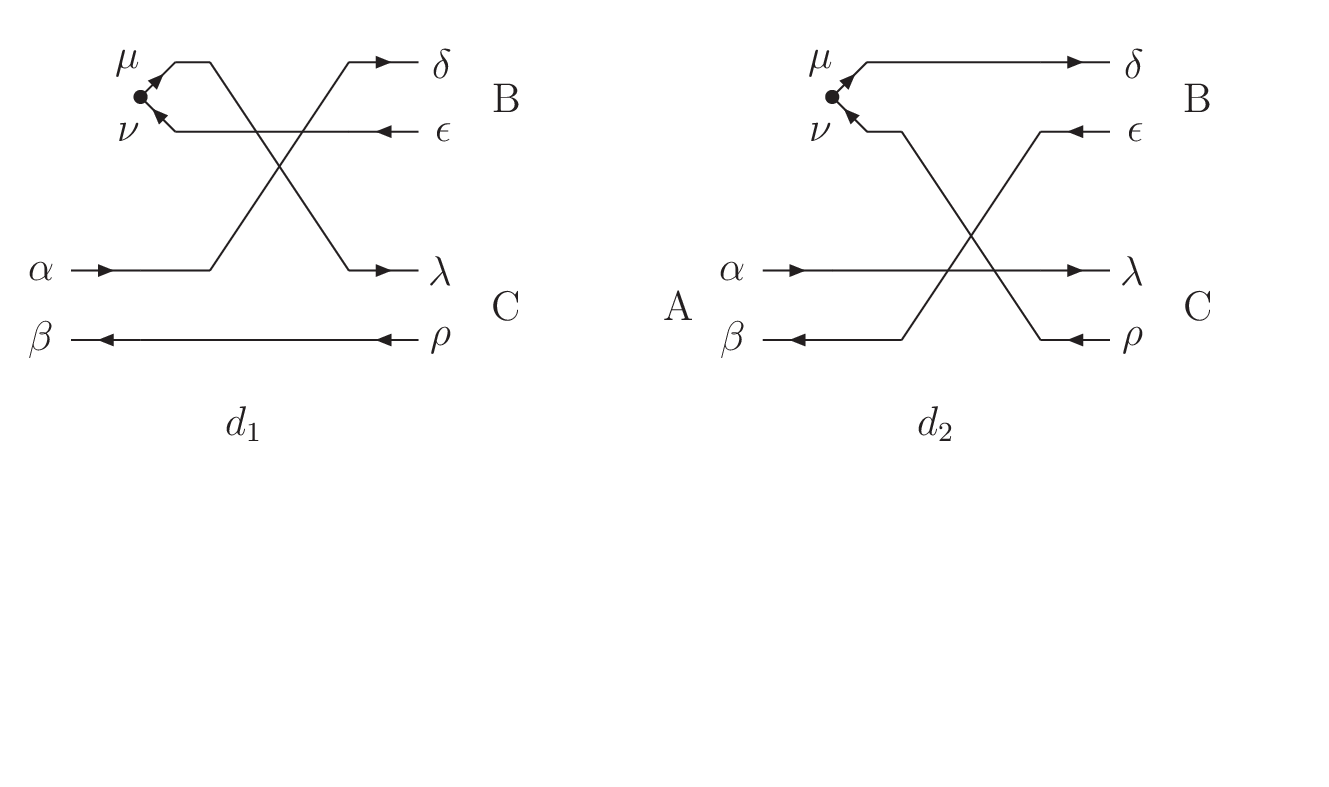,height=3.75cm,width=8.5cm}
\caption{\label{fig:3P0diagrams} Diagrams that can contribute to the decay width
through the $^{3}P_{0}$ model.}
\end{center}
\end{figure}

We are interested on the transition amplitude for the reaction
$(\alpha\beta)_{A} \to (\delta\epsilon)_{B} + (\lambda\rho)_{C}$. The meson $A$
is formed by a quark $\alpha$ and antiquark $\beta$. At some point it is created
a $(\mu\nu)$ quark-antiquark pair. The created $(\mu\nu)$ pair together with the
$(\alpha\beta)$ pair in the original meson regroups in the two outgoing mesons
via a quark rearrangement process. These final mesons are meson $B$ which is
formed by the quark-antiquark pair $(\delta\epsilon)$ and meson $C$ with
$(\lambda\rho)$ quark-antiquark pair.

We work in the center-of-mass reference system of meson $A$, thus we have
$\vec{K}_{A}=\vec{K}_{0}=0$ with $\vec{K}_{A}$ and $\vec{K}_{0}$ the total
momentum of meson $A$ and of the system $BC$ with respect to a given reference
system. We can factorize the matrix element as follow
\begin{equation}
\left\langle BC|T|A\right\rangle =
\delta^{(3)}(\vec{K}_{0}) \mathcal{M}_{A\rightarrow BC}.
\end{equation}

The initial state in second quantization is
\begin{equation}
\left.|A\right\rangle
=\int d^{3}p_{\alpha}d^{3}p_{\beta}\delta^{(3)}(\vec{K}_{A}-\vec{P}_{A})
\phi_{A}(\vec{p}_{A})a_{\alpha}^{\dagger}(\vec{p}_{\alpha})b_{\beta}^{\dagger}
(\vec{p}_{\beta})\left.|0\right\rangle,
\label{eq:Istate}
\end{equation}
where $\alpha$ $(\beta)$ are the spin, flavor and color quantum numbers of
the quark (antiquark). The wave function $\phi_{A}(\vec{p}_{A})$ denotes a meson
$A$ in a color singlet with an isospin $I_{A}$ with projection $M_{I_{A}}$, a
total angular momentum $J_{A}$ with projection $M_{A}$, $J_{A}$ is the coupling
of angular momentum $L_{A}$ and spin $S_{A}$. The $\vec{p}_{\alpha}$ and
$\vec{p}_{\beta}$ are the momentum of quark and antiquark, respectively. The
$\vec{P}_{A}$ and $\vec{p}_{A}$ are the total and relative momentum of the
$(\alpha\beta)$ quark-antiquark pair within the meson $A$. The final state is
more complicated than the initial one because it is a two-meson state. It can be
written as
\begin{widetext}
\begin{equation}
\begin{split}
|BC\!\!\left.\right\rangle =& \frac{1}{\sqrt{1+\delta_{BC}}}\int d^{3}K_{B}
d^{3}K_{C}\sum_{m,M_{BC}}\left\langle\right.
\!\!J_{BC}M_{BC}lm|J_{T}M_{T}\!\!\left.\right\rangle\delta^{(3)}(\vec{K}-\vec{K_
{0}})\delta(k-k_{0}) \\ 
& 
\frac{Y_{lm}(\hat{k})}{k}\sum_{M_{B},M_{C},M_{I_{B}},M_{I_{C}}}\left\langle
J_{B}M_{B}J_{C}M_{C}|J_{BC}M_{BC}\right\rangle\left\langle
I_{B}M_{I_{B}}I_{C}M_{I_{C}}|I_{A}M_{I_{A}} \right\rangle \\ 
& 
\int d^{3}p_{\delta}d^{3}p_{\epsilon}d^{3}p_{\lambda}d^{3}p_{\rho}
\delta^{(3)}(\vec{K}_{B}-\vec{P}_{B})\delta^{(3)}(\vec{K}_{C}-\vec{P}_{C}) \\ 
&
\phi_{B}(\vec{p}_{B})\phi_{C}(\vec{p}_{C})a_{\delta}^{\dagger}(\vec{p}_{\delta}
)b_{\epsilon}^{\dagger}(\vec{p}_{\epsilon})a_{\lambda}^{\dagger}(\vec{p}_{
\lambda})b_{\rho}^{\dagger}(\vec{p}_{\rho})\left.|0\right\rangle,
\label{eq:Fstate}
\end{split}
\end{equation}
\end{widetext}
where we have followed the notation of meson $A$ for the mesons $B$ and $C$. We
assume that the final state of mesons $B$ and $C$ is a spherical wave with
angular momentum $l$. The relative and total momentum of mesons $B$ and $C$ are
$\vec{k}_{0}$ and $\vec{K}_{0}$. The total spin $J_{BC}$ is obtained coupling
the total angular momentum of mesons $B$ and $C$, and $J_{T}$ is the coupling of
$J_{BC}$ and $l$.

The $^{3}P_{0}$ model takes into account only diagrams in which the $(\mu\nu)$
quark-antiquark pair separates into different final mesons. This was originally
motivated by the experiment and it is known as the Okubo-Zweig-Iizuka
(OZI)-rule~\cite{okubo1963,zweigcern2,iizuka1966systematics} which tells us that
the disconnected diagrams are more suppressed than the connected ones. The
diagrams that can contribute to the decay width through the $^{3}P_{0}$ model
are shown in Fig.~\ref{fig:3P0diagrams}.

\subsection{Decay width}

The total width is the sum over the partial widths characterized by the quantum 
numbers $J_{BC}$ and $l$
\begin{equation}
\Gamma_{A\rightarrow BC}=\sum_{J_{BC},l}\Gamma_{A\rightarrow BC}(J_{BC},l),
\end{equation}
where
\begin{equation}
\Gamma_{A\rightarrow BC}(J_{BC},l)=2\pi\int
dk_{0}\delta(E_{A}-E_{BC})|\mathcal{M}_{A\rightarrow BC}(k_{0})|^{2}.
\label{eq:gammaDelta}
\end{equation}

We use relativistic phase space, so
\begin{equation}
\begin{split}
\Gamma_{A\rightarrow 
BC}(J_{BC},l)=2\pi\frac{E_{B}(k_{0})E_{C}(k_{0})}{m_{A}k_{0}}|\mathcal{M}_{
A\rightarrow BC}(k_{0})|^{2},
\end{split}
\end{equation}
where
\begin{equation}
k_{0}=\frac{\sqrt{[m_{A}^{2}-(m_{B}-m_{C})^{2}][m_{A}^{2}-(m_{B}+m_{C})^{2}]}}{
2m_{A}},
\end{equation}
is the on-shell relative momentum of mesons $B$ and $C$.

\section{RUNNING OF THE STRENGTH $\mathbf{\gamma}$ OF THE DECAY INTERACTION}
\label{sec:running}

The strength parameter of the $^{3}P_{0}$ model shows two different type of
dependencies. The first one is the scale with the mass of the pair created
through the relationship with the $g_{s}$ constant, $\gamma=g_{s}/2m$. As in
this work we will study only decays which include the creation of a light quark
pair, this dependence will not be used.

However, if $g_{s}$ is related to fundamental QCD parameters, among them the
strong coupling constant, one expects that $g_{s}$, and hence $\gamma$, depends
on some scale defined by the quark sector.

To elucidate the $\gamma$ dependence on this scale, we calculate through the
$^{3}P_{0}$ model the total strong decay widths of the mesons which belong to
charmed, charmed-strange, hidden charm and hidden bottom sectors.
Table~\ref{tab:fitwidths} shows the experimental data taken for the fit.

\begin{table}[!t]
\begin{center}
\begin{tabular}{cccccccc}
\hline
\hline
Meson & I & J & P & C & Mass (MeV) & $\Gamma_{\rm Exp.}$ (MeV) & \\
\hline
\tstrut
$D_{1}(2420)^{\pm}$ & $1/2$ & $1$ & $+1$ & - & $2423.4\pm3.1$ & $25\pm6$
& \cite{PDG2010} \\
$D_{2}^{\ast}(2460)^{\pm}$ & $1/2$ & $2$ & $+1$ & - & $2460.1\pm4.4$ &
$37\pm6$ & \cite{PDG2010} \\[1ex]
$D_{s1}(2536)^{\pm}$ & $0$ & $1$ & $+1$ & - & $2535.12\pm0.25$ &
$1.03\pm0.13$ & \cite{aubert2006precision} \\
$D_{s2}^{\ast}(2575)^{\pm}$ & $0$ & $2$ & $+1$ & - & $2572.6\pm0.9$ &
$20\pm5$ & \cite{PDG2010} \\[1ex]
$\psi(3770)$ & $0$ & $1$ & $-1$ & $-1$ & $3775.2\pm1.7$ & $27.6\pm1.0$ &
\cite{PDG2010} \\
$\Upsilon(4S)$ & $0$ & $1$ & $-1$ & $-1$ & $10579.4\pm1.2$ & $20.5\pm2.5$ &
\cite{PDG2010} \\
\hline
\hline
\end{tabular}
\caption{\label{tab:fitwidths} Meson decay widths which have been taken into
account in the fit of the scale-dependent strength, $\gamma$. Some properties of
these mesons are also shown.}
\end{center}
\end{table}

\begin{figure}[!t]
\begin{center}
\epsfig{figure=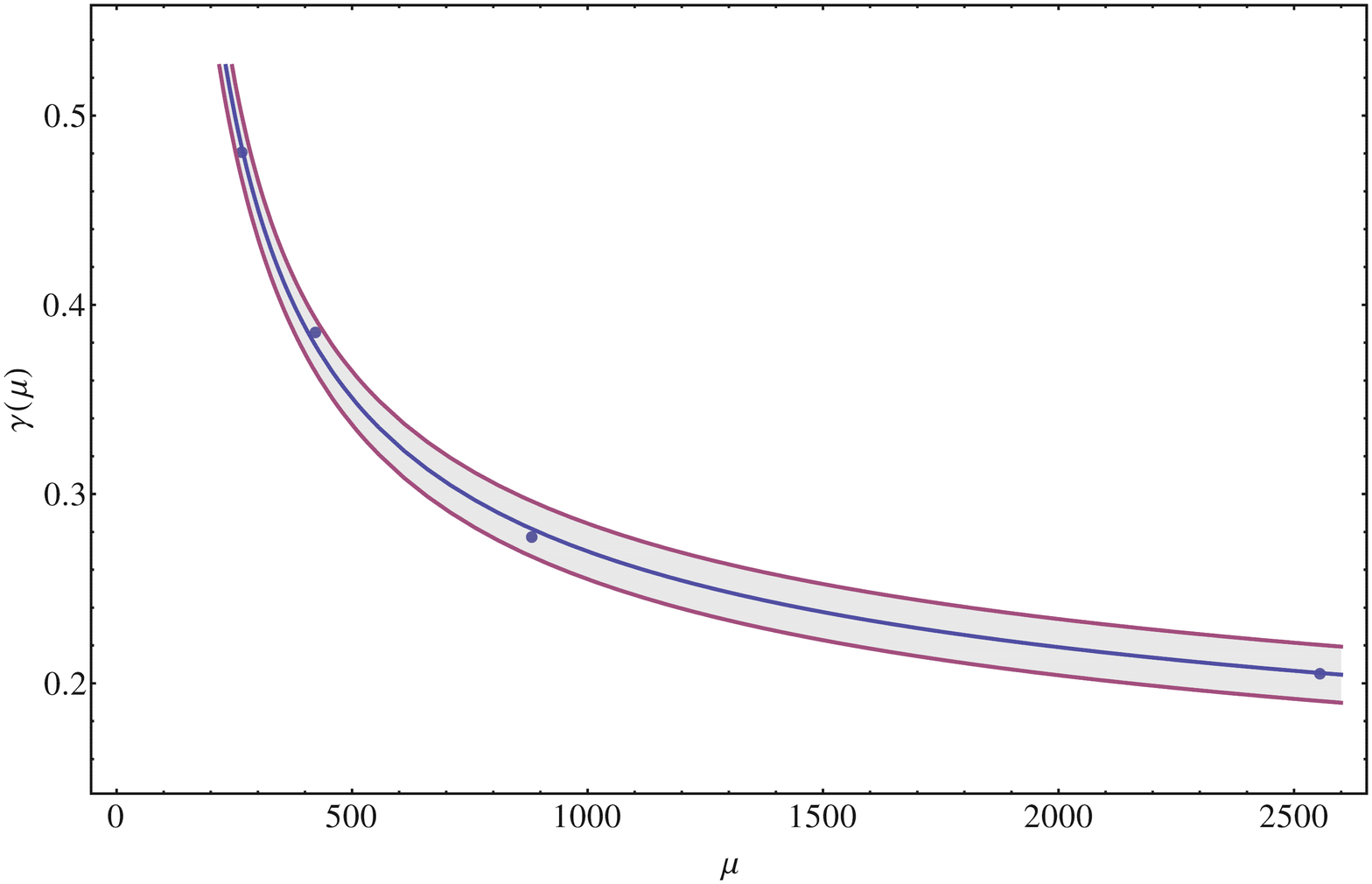,height=6.0cm,width=8.0cm}
\caption {\label{fig:fitgamma} The scale-dependent strength, $\gamma$, in
function of the reduced mass of the $q\bar{q}$ pair of the decaying meson,
$\mu$. The data points are the value of $\gamma$ needed to reproduce the meson
decay widths shown in Table~\ref{tab:fitwidths}. The solid line is the fit and
the shaded area is the confidence interval with $90\%$ confidence level.}
\end{center}
\end{figure}

\begin{table*}[!t]
\begin{center}
\begin{tabular}{lccccccccccccc}
\hline
\hline
\multicolumn{2}{c}{} & \multicolumn{3}{c}{Light mesons} & & 
\multicolumn{4}{c}{Heavy-light mesons} & & \multicolumn{3}{c}{Heavy mesons} \\
& & $(n\bar{n})$ & $(n\bar{s})$ & $(s\bar{s})$ & & $(n\bar{c})$ & $(s\bar{c})$ &
$(n\bar{b})$ & $(s\bar{b})$ & & $(c\bar{c})$ & $(c\bar{b})$ & $(b\bar{b})$ \\
\hline
$\mu$ & & $156.5$ & $200.1$ & $277.5$ & & $265.8$ & $422.1$ & $294.9$ &
$500.6$ & & $881.5$ & $1310.8$ & $2555.0$ \\
$\gamma$ & & $0.707$ & $0.582$ & $0.471$ & & $0.483$ & $0.379$ & $0.455$ &
$0.351$ & & $0.282$ & $0.247$ & $0.205$ \\
\hline
\hline
\end{tabular}
\caption{\label{tab:gammavalues} Values of the scale-dependent strength
$\gamma$ in the different quark sectors following Eq.~(\ref{eq:fitgamma}). The
reduced mass of the $q\bar{q}$ pair in the decaying meson $\mu$ is given in
MeV.}
\end{center}
\end{table*}

In the case of the charmed and charmed-strange sectors, we have considered the
total decay widths of the mesons which belong to the $j_{q}^{P}=3/2^{+}$
doublet predicted by heavy quark symmetry. The reason is that any quark model
predicts the doublet $j_{q}^{P}=\frac{3}{2}^{+}$ in reasonable agreement with
the experiment. Focusing on the $2^{+}$ meson there are no doubts about its
nature and wave function composition. Moreover, in the infinite heavy quark mass
limit these states are narrow, and so we expect that the resonance parameters
are better determined than other states of the same sector. 

For charmonium and bottomonium mesons, we have considered that the best
experimental measurement of total decay widths is that of the state immediately
above the open-flavor sector. This means the total decay width of the
$\psi(3770)$ and $\Upsilon(4S)$ states, respectively.

The decay of $\psi(3770)$ into the $DD$ channel has been widely studied. This
channel is the only open threshold for $\psi(3770)$ and therefore its total
width should be given almost by the decay into $DD$. However, during the last
years this was not the case and the non-$DD$ contribution to the total decay
width was large, $~15\%$. Now, Ref.~\cite{PDG2010} provides a branching
fraction of ${\cal B}(\psi(3770)\to DD)=(93^{+8}_{-9})\%$, which is more
compatible with the theoretical expectations.

The $\Upsilon(4S)$ state is the first one in the bottomonium sector that decays
into a pair of $B$ mesons. In fact, the $\Upsilon(4S)$ resonance decays in
almost $100\%$ of cases to a $BB$ pair, and this feature is exploited by the
$B$-factories to become an important source of data on heavy hadrons in the last
years.

Once the experimental data have been established, we propose a scale-dependent
strength $\gamma$, given by
\begin{equation}
\gamma(\mu) = \frac{\gamma_{0}}{\log\left(\frac{\mu}{\mu_{0}}\right)},
\label{eq:fitgamma}
\end{equation}
where $\mu$ is the reduced mass of the quark-antiquark in the decaying meson
and, $\gamma_{0}=0.81\pm0.02$ and $\mu_{0}=(49.84\pm2.58)\,{\rm MeV}$ are
parameters determined by a global fit of the total decay widths mentioned above.

Fig.~\ref{fig:fitgamma} shows the scale-dependent strength $\gamma$ as a
function of the reduced mass of the decaying meson $\mu$. The data points are
the value of $\gamma$ needed to reproduce the meson decay widths shown in
Table~\ref{tab:fitwidths}. The solid line is the fit and the shaded area is
the confidence interval with $90\%$ confidence level.

For completeness, we show in Table~\ref{tab:gammavalues} the values of the
scale-dependent strength $\gamma$ in the different flavor sectors following
Eq.~(\ref{eq:fitgamma}).

\begin{table*}[!t]
\begin{center}
\begin{tabular}{cccccccccc}
\hline
\hline
Meson & I & J & P & C & n & Mass (MeV) & $\Gamma_{\rm Exp.}$
(MeV)~\cite{PDG2010} & $\Gamma_{\rm The.}$ (MeV) \\
\hline
\tstrut
$D^{\ast}(2010)^{\pm}$ & $0.5$ & $1$ & $-1$ & - & $1$ & $2010.25\pm0.14$ &
$0.096\pm0.022$ & $0.036$ \\
$D_{0}^{\ast}(2400)^{\pm}$ & $0.5$ & $0$ & $+1$ & - & $1$ & $2403\pm38$ &
$283\pm42$ & $212.01$ \\
$D_{1}(2420)^{\pm}$ & $0.5$ & $1$ & $+1$ & - & $1$ & $2423.4\pm3.1$ & $25\pm6$ &
$25.27$ \\
$D_{1}(2430)^{0}$ & $0.5$ & $1$ & $+1$ & - & $2$ & $2427\pm36$ & $384\pm150$ &
$229.12$ \\
$D_{2}^{\ast}(2460)^{\pm}$ & $0.5$ & $2$ & $+1$ & - & $1$ & $2460.1\pm4.4$ &
$37\pm6$ & $64.07$ \\
$D(2550)^{0}$ & $0.5$ & $0$ & $-1$ & - & $2$ & $2539.4\pm8.2$ & $130\pm18$ &
$132.07$ \\
$D^{\ast}(2600)^{0}$ & $0.5$ & $1$ & $-1$ & - & $2$ & $2608.7\pm3.5$ & $93\pm14$
& $96.91$ \\
$D_{J}(2750)^{0}$ & $0.5$ & $\left[\begin{matrix} 2 \\ 3 \end{matrix}\right]$ &
$-1$ & - & $1$ & $2752.4\pm3.2$ & $71\pm13$ & $\left[\begin{matrix} 229.86 \\
107.64 \end{matrix}\right]$ \\
$D_{J}^{\ast}(2760)^{0}$ & $0.5$ & $1$ & $-1$ & - & $3$ & $2763.3\pm3.3$ &
$60.9\pm6.2$ & $338.63$ \\
\hline
\tstrut
$D_{s1}(2536)^{\pm}$ & $0$ & $1$ & $+1$ & - & $1$ & $2535.12\pm0.25$ &
$1.03\pm0.13$~\cite{aubert2006precision} & $0.99$ \\
$D_{s2}^{\ast}(2575)^{\pm}$ & $0$ & $2$ & $+1$ & - & $1$ & $2572.6\pm0.9$ &
$20\pm5$ & $18.67$ \\
$D_{s1}^{\ast}(2710)^{\pm}$ & $0$ & $1$ & $-1$ & - & $2$ & $2710\pm14$ &
$149\pm65$ & $170.76$ \\
$D_{sJ}^{\ast}(2860)^{\pm}$ & $0$ & $\left[\begin{matrix} 1 \\ 3
\end{matrix}\right]$ & $-1$ & - & $\left[\begin{matrix} 3 \\ 1
\end{matrix}\right]$ & $2862\pm6$ & $48\pm7$ & $\left[\begin{matrix} 153.19 \\
85.12 \end{matrix}\right]$ \\
$D_{sJ}(3040)^{\pm}$ & $0$ & $1$ & $+1$ & - & $\left[\begin{matrix} 3 \\ 4
\end{matrix}\right]$ & $3044\pm31$ & $239\pm71$ & $\left[\begin{matrix} 301.52
\\ 432.54 \end{matrix}\right]$ \\[2ex]
\hline
\tstrut
$\psi(3770)$ & $0$ & $1$ & $-1$ & $-1$ & $3$ & $3775.2\pm1.7$ & $27.6\pm1.0$ &
$26.47$ \\
$\psi(4040)$ & $0$ & $1$ & $-1$ & $-1$ & $4$ & $4039\pm1$ & $80\pm10$ & $111.27$
\\
$\psi(4160)$ & $0$ & $1$ & $-1$ & $-1$ & $5$ & $4153\pm3$ & $103\pm8$ & $115.95$
\\
$X(4360)$ & $0$ & $1$ & $-1$ & $-1$ & $6$ & $4361\pm9$ & $74\pm18$ & $113.92$ \\
$\psi(4415)$ & $0$ & $1$ & $-1$ & $-1$ & $7$ & $4421\pm4$ & $62\pm20$ & $159.02$
\\
$X(4640)$ & $0$ & $1$ & $-1$ & $-1$ & $8$ & $4634\pm8$ & $92\pm52$ & $206.37$ \\
$X(4660)$ & $0$ & $1$ & $-1$ & $-1$ & $9$ & $4664\pm11$ & $48\pm15$ & $135.06$
\\
\hline
\tstrut
$\Upsilon(4S)$ & $0$ & $1$ & $-1$ & $-1$ & $6$ & $10579.4\pm1.2$ & $20.5\pm2.5$
& $20.59$ \\
$\Upsilon(10860)$ & $0$ & $1$ & $-1$ & $-1$ & $8$ & $10865\pm8$ & $55\pm28$ &
$27.89$ \\
$\Upsilon(11020)$ & $0$ & $1$ & $-1$ & $-1$ & $10$ & $11019\pm8$ & $79\pm16$ &
$79.16$ \\
\hline
\hline
\end{tabular}
\caption{\label{tab:totalwidths} Strong total decay widths calculated through
the $^{3}P_{0}$ model of the mesons which belong to charmed, charmed-strange,
hidden charm and hidden bottom sectors. The value of the parameter $\gamma$ in
every sector is given by Eq.~(\ref{eq:fitgamma}).}
\end{center}
\end{table*}

\section{RESULTS}
\label{sec:results}

Table~\ref{tab:totalwidths} shows our results for the total strong decay widths
of the mesons which belong to charmed, charmed-strange, hidden charm and hidden
bottom sectors. In the case of mesons containing a single $c$-quark, we have
considered the newly observed charmed mesons $D(2550)$, $D^{\ast}(2600)$,
$D_{J}(2750)$ and $D_{J}^{\ast}(2760)$, and charmed-strange mesons
$D_{s1}^{\ast}(2710)$, $D_{sJ}^{\ast}(2860)$ and $D_{sJ}(3040)$. Our model
predicts as naive $c\bar{c}$ states the $X(4360)$, $X(4640)$ and $X(4660)$
mesons, they are also included in the study of the charmonium sector. The
bottomonium states are the usual ones above the $BB$ threshold.

We get a quite reasonable global description of the total decay widths. The
detailed analysis of the decay modes of every resonance is beyond the scope of
this work, whose main goal is to establish a scale dependence for $\gamma$.
However, let us comment in more detail each sector discussing briefly the most
significant aspects.

The results predicted by the $^{3}P_{0}$ model for the well established charmed
mesons are in good agreement with the experimental data except for one case,
the total decay width of the $D^{\ast}$ meson. The $D^{\ast}$ decays only into
$D\pi$ channel via strong interaction and it is assumed that the total decay
width is given mainly by this decay mode. However, the disagreement may be due
to the very small available phase space which enhances possible effects of the
final-state interactions.

In Ref.~\cite{PhysRevD.82.111101} the BaBar Collaboration reported the new
charmed states $D(2550)$, $D^{\ast}(2600)$, $D_{J}(2750)$ and
$D_{J}^{\ast}(2760)$ in inclusive $e^{+}e^{-}$ collisions. The $J^{P}=0^{-}$ is
the most plausible assignment for the $D(2550)$ meson, the total width predicted
by the $^{3}P_{0}$ model with this assignment is in very good agreement with the
experimental data. The helicity-angle distribution of $D^{\ast}(2600)$ is found
to be consistent with $J^{P}=1^{-}$. Moreover, its mass makes it the perfect
candidate to be the spin partner of the $D(2550)$ meson. Our prediction of the
total decay width as the $2^{3}S_{1}$ state agrees again with the data. There is
a strong discussion in the literature about the possible quantum numbers that
could have the mesons $D_{J}(2750)$ and $D_{J}^{\ast}(2760)$ providing a wide
range of assignments. The total strong decay widths of these mesons have been
calculated attending to the most plausible assignment coming from our model.
While there seems to be a consistent assignment to the $D_{J}(2750)$ meson, it
is not the case for the $D_{J}^{\ast}(2760)$ one.

In Ref.~\cite{PhysRevD.80.054017} we have considered the coupling between the
$1^{+}$ $c\bar{s}$ states and a tetraquark, finding that the $J^{P}=1^{+}$
$D_{s1}(2460)$ has an important non-$q\bar{q}$ contribution whereas the
$J^{P}=1^{+}$ $D_{s1}(2536)$ is almost a pure $q\bar{q}$ state. The presence of
non-$q\bar{q}$ degrees of freedom in the $J^{P}=1^{+}$ charmed-strange meson
sector enhances the $j_{q}=3/2$ component of the $D_{s1}(2536)$. This wave
function explains most of the experimental data, as shown in
Refs.~\cite{PhysRevD.80.054017,segovia2011semileptonic}, and it is the one we
use here.

\begin{table*}[!t]
\begin{center}
\begin{tabular}{lll}
\hline
\hline
Decay mode & Theory & Experiment \\
\hline
$f_{0}(600)\to \pi\pi$ & $\Gamma_{\pi\pi}=(224-651)\,{\rm MeV}$ &
$\Gamma_{\rm tot}=(250-500)\,{\rm MeV}$ \\
$h_{1}(1170)\to \rho\pi$ & $\Gamma_{\rho\pi}=619\,{\rm MeV}$ &
$\Gamma_{\rm tot}=(360\pm40)\,{\rm MeV}$ \\
$f_{2}(1270)\to \pi\pi$ & $\Gamma_{\pi\pi}=315\,{\rm MeV}$ &
$\Gamma_{\pi\pi}=(156.9^{+4.0}_{-1.2})\,{\rm MeV}$ \\
\hline
$\rho\to \pi\pi$ & $\Gamma_{\pi\pi}=160\,{\rm MeV}$ &
$\Gamma_{\pi\pi}=(148.1\pm0.6)\,{\rm MeV}$ \\
$b_{1}(1235)\to \omega\pi$ & $\Gamma_{\omega\pi}=158\,{\rm MeV}$ & $\Gamma_{\rm
tot}=(142\pm9)\,{\rm MeV}$ \\
& ${\cal B}((\omega\pi)_{\rm S-wave})=0.76$ & - \\
& ${\cal B}((\omega\pi)_{\rm D-wave})=0.24$ & - \\
& ${\cal B}((\omega\pi)_{\rm D-wave})/{\cal B}((\omega\pi)_{\rm
S-wave})=0.32$ & ${\cal B}((\omega\pi)_{\rm D-wave})/{\cal
B}((\omega\pi)_{\rm S-wave})=0.277\pm0.027$ \\[1.5ex]
$a_{1}(1260)\to \rho\pi$ & $\Gamma_{\rho\pi}=837\,{\rm MeV}$ &
$\Gamma_{\rm tot}=(250-600)\,{\rm MeV}$ \\
& ${\cal B}((\rho\pi)_{\rm S-wave})=0.91$ & ${\cal B}((\rho\pi)_{\rm
S-wave})=0.6019$ \\
& ${\cal B}((\rho\pi)_{\rm D-wave})=0.09$ & ${\cal B}((\rho\pi)_{\rm
D-wave})=0.013\pm0.0060\pm0.0022$ \\[1.5ex]
$a_{2}(1320)\to \rho\pi$ & $\Gamma_{\rho\pi}=255\,{\rm MeV}$ & - \\
$a_{2}(1320)\to \eta\pi$ & $\Gamma_{\eta\pi}=69\,{\rm MeV}$ &
$\Gamma_{\eta\pi}=(18.5\pm3.0)\,{\rm MeV}$ \\
$a_{2}(1320)\to \eta'\pi$ & $\Gamma_{\eta'\pi}=12\,{\rm MeV}$ &
$\Gamma_{\eta'\pi}=(0.59\pm0.10)\,{\rm MeV}$ \\
\hline
\hline
\end{tabular}
\caption{\label{tab:SDlight} Some strong decay observables of light mesons
calculated through the $^{3}P_{0}$ model with the value of the parameter
$\gamma$ given by Eq.~(\ref{eq:fitgamma}). The theoretical range on the total
decay width of the $f_{0}(600)$ is obtained moving the mass of the $f_{0}(600)$
in its experimental range.}
\end{center}
\end{table*}

Two new charmed-strange resonances, the $D_{s1}^{\ast}(2710)$ and
$D_{sJ}^{\ast}(2860)$, have been observed by the BaBar Collaboration in both
$DK$ and $D^{\ast}K$ channels~\cite{PhysRevD.80.092003}. In the $D^{\ast}K$
channel, the BaBar Collaboration have also found evidence for the
$D_{sJ}(3040)$, but there is no signal in the $DK$ channel. It is commonly
believed that the $D_{s1}^{\ast}(2710)$ is the first excitation of
the $D_{s}^{\ast}$ meson. With this assignment, the prediction of the
$^{3}P_{0}$ model is in agreement with the experimental data. In
Table~\ref{tab:totalwidths} we show the total strong decay width of the
$D_{sJ}^{\ast}(2860)$ as the third excitation of the $1^{-}$ meson and as the
ground state of the $3^{-}$ meson. The comparison between experimental data and
our results favors the $n\,J^{P}=1\,3^{-}$ assignment. The mean $2P$ multiplet
mass is predicted in our model to be near the mass of the $D_{sJ}(3040)$
resonance. The only decay mode in which $D_{sJ}(3040)$ has been seen until now
is the $D^{\ast}K$, and so the most possible assignment is that the
$D_{sJ}(3040)$ meson being the next excitation in the $1^{+}$ channel.
Table~\ref{tab:totalwidths} shows our prediction of the $D_{sJ}(3040)$ decay
width as the $nJ^{P}=3\,1^{+}$ or $4\,1^{+}$ state. Both are large but
compatible with the experimental data.

From an experimental point of view there are a few data in the open-charm decays
of the $1^{--}$ $c\bar{c}$ resonances. The main experimental data are the
resonance parameters, mass and total decay width, of the excited $\psi$ states
fitting the $R$ value measured in the relevant energy region. One can see that
the general trend of the total decay widths is well reproduced. There are two
particular cases in which the theoretical results exceed the experimental one.
The first case is the $\psi(4415)$ where we predict a total width of
$159\,{\rm MeV}$, while the PDG~\cite{PDG2010} average value is
$62\pm20\,{\rm MeV}$. However one should mention that the experimental data are
clustered around two values ($\sim\!100\,{\rm MeV}$ and $\sim\!50\,{\rm MeV}$)
corresponding the lower one to very old measurements. If we compare our result
with the recent experimental data reported by Seth {\it et
al.}~\cite{PhysRevD.72.017501} ($\Gamma=119\pm16\,{\rm MeV}$), they are more
compatible. The second result which disagrees with the experimental data is the
corresponding to the pair of states in the vicinity of $4.6\,{\rm GeV}$. Both
widths are larger than the experimental results. The smallest total width of the
$X(4660)$ favors the $4^{3}D_{1}$ option for this state although interference
between the two states can be the origin of the small experimental width.

We obtain a very good agreement between experimental and theoretical total decay
widths in the bottomonium sector. The most significant disagreement is found for
the $\Upsilon(5S)$ state, note however the large error in the experimental data.

Finally, one may wonder what happens in other sectors in which the fit has not
been carried out. The question may be more obvious in the light quark sector
where the $^{3}P_{0}$ model has been extensively used with different values of
$\gamma$. We do not expect to accurately describe the strong decays of light
mesons, but it would be an achievement of the parametrization to obtain light
meson widths on the order of the experimental ones. Table~\ref{tab:SDlight}
shows the theoretical results for some decay modes in the light quark sector and
compares with the available experimental data. There are cases in which the
agreement is evident, but others do not quite agree with the data. We obtain
always the order of magnitude of the total decay widths.

Another sector not included in the fit is the open-bottom sector. Although the 
experimental data are scarce, we can focus on the orbitally excited $B$ mesons
which has been recently measured by the D0 and CDF Collaborations. There are two
well established states, the $B_{1}(5721)$ and $B_{2}^{\ast}(5747)$ mesons. The
CDF Collaboration has reported the width of the $B_{2}^{\ast}(5747)$ and from
this one, the width of the $B_{1}(5721)$ can be estimated using the result of
Ref.~\cite{PhysRevD.53.231}. In Table~\ref{tab:Bmesons} we show the predicted
widths for these states. One can see a good agreement with the experimental data
despite of the fact that the expression for the $\gamma$ running has not been
fitted in this sector. Moreover, as the reduced mass in the $B$ meson is closer
to that of the light meson than to that of the heavy meson, this data cannot be
reproduced if we use a $\gamma$ value which fits the bottomonium decays.
Although it is independent of $\gamma$ the ratio ${\cal R}=\frac{{\cal
B}(B^{\ast}_{2}\to B^{\ast}\pi)}{{\cal B}( B^{\ast}_{2}\to
B^{(\ast)}\pi)}=0.475\pm0.095\pm0.069$ gives $0.49$, in excellent agreement with
the data.

\begin{table}[!t]
\begin{center}
\begin{tabular}{cccc}
\hline
\hline
Meson & Decay mode & $\Gamma_{\rm The.}$ (MeV) &  $\Gamma_{\rm Exp.}$ (MeV) \\
\hline
$B_{1}(5721)^{0}$ & $B^{\ast0} \pi^0$ & $6.8$  & \\
                  & $B^{\ast+} \pi^-$ & $13.6$ & \\
                  & total             & $20.4$ & $20.4\pm 4.5\pm 9.6$ \\
$B_2(5747)^0$ & $B^{0}\pi^{0}$     & $5.7$  & \\
              & $B^{+}\pi^{-}$     & $11.3$ & \\
              & $B^{\ast0}\pi^{0}$ & $5.3$  & \\
              & $B^{\ast+}\pi^{-}$ & $10.6$ & \\
              & total              & $32.9$ & $22.7\pm 5.0\pm 10.7$ \\
\hline
\hline
\end{tabular}
\caption{\label{tab:Bmesons} Open-flavor strong decay widths, in MeV, of the
$B_{1}(5721)$ and $B_{2}^{\ast}(5747)$ mesons calculated through the
$^{3}P_{0}$ model with the value of the parameter $\gamma$ given by
Eq.~(\ref{eq:fitgamma}).}
\end{center}
\end{table}

\section{CONCLUSIONS}
\label{sec:conclusions}

We propose a scale-dependent strength $\gamma$ of the phenomenological
$^{3}P_{0}$ model as a function of the reduced mass of the quark-antiquark pair
of the decaying meson to achieve a global description of the meson strong
decays. The dependence of $\gamma$ has been taken as logarithmically in the
reduced mass.

To do that we have performed a calculation of the total strong decay widths of
the mesons which belong to charmed, charmed-strange, hidden charm and hidden
bottom sectors. The wave functions for the mesons involved in the open-flavor
strong decays are given by the potential model described in
Ref.~\cite{vijande2005constituent} which has been successfully applied to
hadron phenomenology and reactions.

The results predicted by the $^{3}P_{0}$ model with the suggested running of the
$\gamma$ parameter are in a global agreement with the experimental data, being
remarkable in most of the cases studied. For mesons containing a single
$c$-quark, we have considered the newly observed charmed mesons ($D(2550)$,
$D^{\ast}(2600)$, $D_{J}(2750)$ and $D_{J}^{\ast}(2760)$) and charmed-strange
mesons ($D_{s1}^{\ast}(2710)$, $D_{sJ}^{\ast}(2860)$ and $D_{sJ}(3040)$). In the
charmonium sector, possible $XYZ$ assignments have been considered. We obtain
good agreement between theoretical and experimental decay widths for the
$\Upsilon$ states which are above the open-bottom threshold.

For completeness, we provide some predictions in other sectors in which the fit
has not been carried out. The light quark sector shows that our parametrization
is not so far of the real picture. The predictions in the open-bottom sector
where the reduced mass in the $B$ meson is closer to that of the light meson
are in very good agreement with the available experimental data.

\begin{acknowledgments}

This work has been partially funded by Ministerio de Ciencia y Tecnolog\'ia
under Contract No. FPA2010-21750-C02-02, by the European Community-Research
Infrastructure Integrating Activity 'Study of Strongly Interacting Matter'
(HadronPhysics2 Grant No. 227431) and by the Spanish Ingenio-Consolider 2010
Program CPAN (CSD2007-00042).

\end{acknowledgments}


\bibliographystyle{apsrev}
\bibliography{running3P0}

\end{document}